\documentclass[acmsmall]{acmart}

\usepackage{colortbl} 
\usepackage{xcolor} 
\usepackage{array} 
\usepackage{multirow} 
\usepackage{geometry} 
\usepackage{natbib}
\usepackage{soul}

\usepackage[normalem]{ulem}

\usepackage{adjustbox}
\newcolumntype{P}[1]{>{\raggedright\arraybackslash}p{#1}}

\definecolor{darkgreen}{rgb}{0.0, 0.5, 0.0}

\usepackage[most]{tcolorbox} 

\newtcolorbox{bubble}[1]{colback=black!5!white, 
  colframe=black!75!black,
  fonttitle=\bfseries,
  title=#1,
  width=\linewidth,
  boxsep=1pt,
  top=1pt,
  bottom=1pt}

\let\svthefootnote\thefootnote
\newcommand\blankfootnote[1]{%
  \let\thefootnote\relax\footnotetext{#1}%
  \let\thefootnote\svthefootnote%
}


\begin{document}

\title[Charting Uncertain Waters: A Socio-Technical Roadmap for Sustaining Open Source Communities in the Age of GenAI.]{Charting Uncertain Waters: A Socio-Technical Roadmap for Sustaining Open Source Communities in the Age of GenAI}

\author{Zixuan Feng}
\email{fengzi@oregonstate.edu}
\orcid{0000-0001-9163-6853}
\affiliation{%
  \institution{Oregon State University}
  \country{USA}
}

\author{Reed Milewicz}
\affiliation{%
  \institution{Sandia National Laboratories}
  \country{USA}}
\email{rmilewi@sandia.gov}

\author{Emerson Murphy-Hill}
\affiliation{%
  \institution{Microsoft}
  \country{USA}}
\email{emerson.rex@microsoft.com}

\author{Tyler Menezes}
\affiliation{%
  \institution{CodeDay}
  \country{USA}}
\email{tylermenezes@codeday.org}

\author{Alexander Serebrenik}
\affiliation{%
  \institution{Eindhoven University of Technology}
  \country{Netherlands}}
\email{a.serebrenik@tue.nl}

\author{Igor Steinmacher}
\affiliation{%
  \institution{Northern Arizona University}
  \country{USA}}
\email{igor.steinmacher@nau.edu}

\author{Anita Sarma}
\affiliation{%
  \institution{Oregon State University}
  \country{USA}}
\email{Anita.Sarma@oregonstate.edu}

\begin{abstract}
Open Source Software (OSS) communities face a wave of uncertainty as Generative AI (GenAI) rapidly transforms how software is created, maintained, and governed.  Without clear frameworks, communities risk being overwhelmed by the complexity and ambiguity introduced by GenAI,  threatening the collaborative ethos that underpins OSS. 
To address this gap, we present a Socio-Technical System (STS)-guided conceptual framework that applies McLuhan’s Tetrad as an analytic lens to articulate how GenAI reshapes the socio-technical dynamics of OSS development. Through a scenario-based exploration across four components of the STS-guided framework—software practices, documentation, community engagement, and governance—we identify plausible socio-technical impacts and outline a corresponding Roadmap for sustaining OSS communities in the Age of GenAI. This Roadmap will enable OSS researchers and practitioners to interpret emerging transformations, anticipate challenges, and design interventions that foster long-term community resilience.
By adopting this framework, OSS leaders and researchers can proactively shape the future of their ecosystems, rather than simply reacting to technological upheaval.
\end{abstract}


\maketitle

\section{Introduction}

Open Source Software (OSS) is a profoundly human enterprise, one which owes its success to a dynamic alliance of numerous communities of contributors worldwide, all working together to build the software that powers the modern world. The immense value of OSS infrastructures to the global economy, which by one estimate weighs in at \$8.8 trillion USD~\cite{hoffmann2024value}, is inseparable from the people creating that software and their values of altruism, freedom, mutual respect, and collaboration~\cite{gerosa2021shifting, yue2025discovering}. With this in mind, this paper explores the future of the OSS ecosystem, its health, and sustainability, with a primary focus on its people and what can be done to empower them to do their best work.

At the forefront of emerging trends and technologies that may impact OSS is Generative AI (GenAI), which is rapidly gaining traction in software engineering by transforming how communities collaborate, create code, and approach problem-solving \cite{jackson2024creativity}. Yet despite its growing influence, the integration of GenAI into OSS ecosystems raises uncertainties around its potential to transform the nature of OSS ecosystems. On one hand, GenAI creates many new possibilities that were previously inaccessible or difficult to achieve. For example, contributors who may have lacked formal training in programming or struggled with complex toolchains can now use GenAI tools to generate, debug, and comprehend code \cite{qiao2025systematic, prather2024widening, qiao2025comprehension}. On the other hand, over-reliance on GenAI also introduces new challenges and risks, such that homogenization of code and problem-solving approaches may reduce opportunities for creative exploration, and weaken interpersonal connections \cite{denny2024computing, mohamed2025impact, alami2025human}.

As with any disruptive technology, GenAI introduces a wave of uncertainty and competing visions. It can empower, automate, and simplify, but also confuse, displace, and homogenize. Simply incorporating AI tools, such as chatbots or code generators, does not automatically improve collaboration or learning within OSS projects \cite{dou2024s, alami2025human}. A recent study has begun to discuss GenAI's impact on developer creativity \cite{jackson2024creativity}; however, less is known about how it will reshape the sustainability and governance of OSS ecosystems. Yet this question is just as, if not more, urgent. Unlike conventional software environments, OSS projects are deeply socio-technical, shaped not only by code and tools but also by values, governance practices, and collaborative relationships \cite{feng2022case}. These human infrastructures are now being reconfigured by GenAI. If communities wait passively for GenAI companies to ``boil the ocean'' by universally solving all challenges, they risk losing control over the very principles that make OSS resilient and inclusive. Instead, communities must grapple with deeper questions: \emph{What ways of working does GenAI enhance, render obsolete, retrieve, or reverse? How do we in OSS make GenAI work for \textit{us}}?

To navigate these uncertainties, this paper adopts a conceptual research approach to explore how GenAI may reconfigure the socio-technical fabric of OSS communities and introduce new tensions, possibilities, and trajectories. We do not present an empirical study in this paper because the landscape of GenAI in OSS is still rapidly evolving, with many of its long-term effects yet to be fully materialized. 
Instead, we apply McLuhan’s Tetrad \cite{mcluhan} as an analytic lens for collaborative scenario development as recommended in the disruptive research playbook \cite{storey2024disruptive}; we use the Socio-Technical Systems (STS)-guided framework \cite{leavitt2013applied, lyytinen2008explaining, ciriello2024emergence, bostrom1977mis, perozzo2022cybersecurity}, to guide our scenario development.

McLuhan’s Tetrad guides the examination of how technological innovations reshape socio-technical systems through four interpretive questions: (1) what they enhance or amplify, (2) what they obsolesce or render obsolete, (3) what they retrieve from past practices, and (4) what they reverse into when pushed to extremes. We then use the STS-guided framework to create interpretative scenarios along the four STS dimensions, tasks, technologies, people, and organization \cite{leavitt2013applied, lyytinen2008explaining, ciriello2024emergence, bostrom1977mis, perozzo2022cybersecurity}, to envision how GenAI may reconfigure OSS practices, governance, and community participation. These scenarios surface emerging risks, possibilities, and trajectories of GenAI usage in OSS.

This paper makes two primary contributions. First, we contribute a set of structured, scenario-based explorations of how GenAI may reshape the OSS socio-technical landscape by systematically unpacking plausible impacts across: software development practices (\textit{tasks}), documentation \& accessibility (\textit{technology}), community engagement \& collaboration (\textit{people}), and project governance (\textit{structures}). Second, it distills these insights into a forward-looking research agenda across both near-term adaptations and far-reaching transformations as OSS communities respond to GenAI integration.

The remainder of this paper is organized as follows. Section II discusses the existing literature on GenAI and its emerging influence on software engineering and OSS. Section III outlines our conceptual scaffold, explaining the scenario-based, vision-driven approach and socio-technical framing we use to construct and structure our discussion. Section IV presents plausible future scenarios that illustrate how GenAI may shape OSS ecosystems, applying McLuhan’s Tetrad \cite{mcluhan} across four STS components to reveal potential impacts and tensions. Finally, Section V discusses forward-looking research directions and future empirical work opportunities as the software engineering community navigates GenAI integration in OSS.

\textbf{Positionality Statement.} Our vision paper reflects insights from a multidisciplinary team of seven researchers, including four from academia and three from industry. Together, our collective expertise lies at the intersection of OSS sustainability, socio-technical systems, and GenAI. The academic authors comprise four software engineering researchers with extensive experience studying OSS ecosystems and socio-technical systems, particularly in areas such as sustainability, inclusivity, mentoring, and contributor onboarding. The industry authors include: one researcher from Microsoft, who specializes in developer experience with AI-based developer tools; another from a national laboratory, who focuses on improving practices, processes, and tools across the software development lifecycle; and a third, who serves as the CEO of the CodeDay mentoring program. This diverse composition equips us with a robust understanding of the technical, social, and governance dimensions of OSS, as well as the transformative potential of GenAI. Our interdisciplinary backgrounds enable us to approach the challenges of GenAI integration in OSS with a balanced perspective, combining theoretical insights with practical expertise.

\section{Setting the Stage}
\label{sec:rw}

We present a brief introduction to recent works on GenAI usage in software development and OSS. GenAI is transforming software development and increasingly influencing OSS communities \cite{sauvola2024future, coutinho2024role, nguyen2023generative}. A 2024 Linux Foundation survey found that 94\% of OSS organizations now use GenAI for tasks such as automation, documentation, and code generation \cite{lawson2024shaping}.

Empirical studies have begun to quantify this impact. On one hand, GenAI tools like GitHub Copilot boosted OSS productivity—raising project-level output by 5.9\%, with increases in individual productivity (2.1\%) and participation (3.4\%) \cite{song2024impact}. On the other hand, \citet{yeverechyahu2024impact} found that GenAI contributions leaned heavily toward maintenance and refinement, with fewer instances of novel feature development.

In addition, studies have identified technical risks and collaboration challenges associated with GenAI in software development. \citet{gao2024current} documented 26 challenges across design, testing, and code review. Others have raised concerns about hallucinations, data privacy, and the difficulty of validating AI-generated outputs \cite{madampe2025we, fan2023large}. At the same time, some tools aim to improve accessibility, such as LLM-powered documentation systems \cite{luo2024repoagent}, while ChatGPT has been used to support collaboration and code review in pull requests \cite{chouchen2024software}.

As GenAI becomes increasingly embedded in OSS workflows, its influence extends beyond technical productivity, with the potential to reshape social dynamics, community norms, and participation structures \cite{feng2025modeling, afroz2025developer}. This is especially relevant in OSS, a socio-technical digital ecosystem defined by decentralized governance, volunteer-driven contributions, and distinct collaboration practices. One of the early socio-technical research directions on GenAI in software engineering is by \citet{hyrynsalmi2024making}, which examines how automation and GenAI influence diversity and inclusion in software development. Although not specific to OSS, their insights provide an early discussion for understanding GenAI’s social implications and motivate extending such inquiry to OSS sustainability.

Moreover, existing works that have also attempted to envision AI-infused future scenarios are \citet{trinkenreich2025get} and \citet{jackson2024creativity}. \citet{trinkenreich2025get}, through a call to action, lays out the pros and cons of using LLM as part of the SE research pipeline. \citet{jackson2024creativity} investigates the potential impacts of GenAI on creativity in software development to propose a research agenda on this topic. Both these works leverage McLuhan’s Tetrad \cite{mcluhan}, based on the guidelines recommended in the Disruptive Research Playbook \cite{storey2024disruptive}.

Following this established methodological approach, we similarly apply McLuhan's Tetrad to explore how GenAI may reshape the socio-technical fabric of OSS communities and their collaborative practices. While \citet{jackson2024creativity} applied McLuhan's Tetrad within a 4P framework (focusing on Product, Process, Person, and Press dimensions), we instead employ an STS-guided analytical framework \cite{leavitt2013applied, lyytinen2008explaining, ciriello2024emergence, bostrom1977mis, perozzo2022cybersecurity}. The 4P framework emphasizes individual and product-centric strategies but neglects the broader social dynamics and technical structures relevant to OSS communities \cite{constantinides2006marketing, rhodes1961analysis, tapp20134ps}. The STS-guided framework explicitly addresses the interdependence of social and technical elements in organizational contexts. Prior socio-technical research demonstrates the value of STS-guided analysis in examining how human collaboration and technical infrastructure interact in software teams, including analyses of project failure, software risk management, and collaborative practices \cite{kolukuluri2023qualitative, persson2009managing, ciriello2024emergence}.

The following section outlines our conceptual approach for analyzing GenAI's impact on OSS communities through the STS-guided McLuhan Tetrad framework.

\section{Conceptual Scaffold and Approach}
\label{sec:conceptualApproach}

To investigate the impacts of GenAI on the sustainability of OSS communities, we conducted a collaborative scenario development exercise with members of our research team, all of whom have experience studying or participating in OSS communities. We applied McLuhan’s Tetrad as an analytic lens \cite{mcluhan}  to create interpretative scenarios, as recommended in the disruptive research playbook by  \citet{storey2024disruptive}, within the four components of the STS-guided framework \cite{leavitt2013applied, lyytinen2008explaining, ciriello2024emergence, bostrom1977mis, perozzo2022cybersecurity}. We use the STS-guided framework since it offers a comprehensive and systematic lens for capturing the interdependencies between social dynamics and technical structures. Next, we paint these plausible futures to help guide community reflection and future empirical work.

\subsection{Tetrad Analysis} 
\textbf{McLuhan Tetrad Analysis: Unpacking GenAI’s Effects.} We draw on the Disruptive Research Playbook \cite{storey2024disruptive}, which was specifically designed to help identify socially relevant software development research questions when studying disruptive technologies through the lens of McLuhan's Tetrad \cite{mcluhan}. The tetrad poses four questions about a technology’s effects: (1) What does it enhance or amplify? (2) What does it obsolesce or render obsolete? (3) What does it retrieve from the past? (4) What does it reverse into when pushed to extremes?

\begin{figure}[!bt]
    \centering
    \includegraphics[width = 0.5\textwidth]{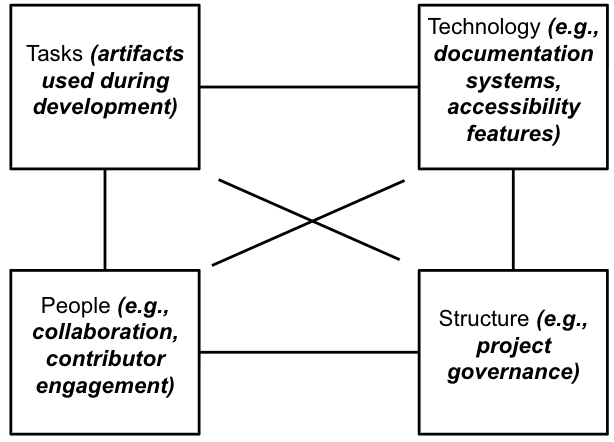}
    \caption{Using the STS-guided framework to understand the impact of GenAI on the sustainability of OSS communities}
    \label{fig:STS}
\end{figure}

We structure the tetrad analysis around the four interrelated components of the STS-guided framework as shown in the Figure \ref{fig:STS}. \emph{Tasks} refers to the activities, processes, and practice sequences through which the system accomplishes its objectives (e.g., contributing code changes, improving code quality, resolving vulnerabilities). \emph{Technology} component encompasses the technical mechanisms, infrastructures, and artifacts that mediate work practices. In OSS, this includes the technical components and socio-technical artifacts that shape how contributors access information, coordinate with others, and accomplish tasks.
\emph{People} include the individuals or groups who interact with the system, including their roles, behaviors, and values (e.g., collaboration, contributor engagement). \emph{Structure} represents the organizational framework, including roles, relationships, and governance mechanisms within the system (e.g., project governance).

These four components are inherently interconnected; a change in one dimension often produces ripple effects across the others \cite{leavitt2013applied, lyytinen2008explaining}. For example, the introduction of a new technology (such as an automated documentation tool) can reshape the nature of tasks performed, alter collaboration patterns among people, and necessitate updates in governance structures. Similarly, changes in project structure or governance can influence who participates (People), how tasks are performed, and which technologies are prioritized or adopted.  This web of interdependencies foregrounds how socio-technical change is not isolated but emerges from the continual interaction and mutual shaping among technical systems, human practices, and organizational arrangements. This integrated perspective is especially well-suited to OSS ecosystems, where rapid technological innovation, distributed collaboration, and evolving governance structures are deeply intertwined. Applying the STS-guided framework enables us to systematically analyze the impact of GenAI on tasks, tools, contributors, and project organization (structures), collectively shaping—and being shaped by—the unique dynamics of OSS development.

\subsection{Constructing Scenarios of GenAI’s Socio-Technical Impact} Building on the STS-guided framework, we used McLuhan’s Tetrad as an analytic lens to structure our scenario development around four socio-technical components. This integrated approach provided both a conceptual scaffold and an organizing structure for how impacts were generated, grouped, and refined during the brainstorming process.

Our scenario development process was conducted through a series of iterative brainstorming sessions over three months. To mitigate common limitations of group brainstorming, such as social loafing and normative pressure \cite{ritter2018facilitate, diehl1987productivity, paulus2007toward}, the first author initially invited each author to independently generate potential GenAI-related impacts based on their own disciplinary expertise. These individual reflections were then synthesized and discussed collaboratively, where the team refined, merged, and categorized ideas into the four Tetrads.

Finally, we validated the synthesized scenarios by inviting two experts who are not co-authors. One of these experts is a veteran open-source researcher and educator with over three decades of experience in software engineering and community development, while another is an active open-source practitioner. During these conversations, the experts discussed how GenAI might enhance, obsolesce, retrieve, or reverse across the four STS components based on their experiences. We did not present our synthesized results during this stage to avoid confirmation bias \cite{nickerson1998confirmation}. Instead, we followed the same process as with other authors, first inviting the experts to share potential GenAI-related impacts from their experience, then prompting reflection on areas they had not raised, guided by our synthesized results (e.g., How might GenAI reshape testing practices? What might be enhanced, obsolesced, retrieved, or reversed?). The two experts did not introduce new results beyond our synthesis; instead, they validated our findings by offering higher-level articulations or concrete examples of the same underlying dynamics. This expert validation aligns with established practices in which knowledgeable domain experts evaluate the plausibility, completeness, and conceptual soundness of a proposed model or synthesis \cite{patton2014qualitative, jaccard2019theory}.

\section{The future of OSS in the era of GenAI}
\label{sec:vision}

Here, we present plausible future scenarios that envision how GenAI might influence the OSS landscape and how these imagined trajectories could reconfigure tasks, technologies, people, and structures within OSS communities based on the STS-guided McLuhan Tetrad framework. These scenarios are interpretive; they do not forecast exact outcomes but instead surface possible directions and tensions to guide reflection and future inquiry.

Our brainstorming sessions were guided by STS components to address both shorter-term, operational dynamics (e.g., immediate changes in coding, documentation, and collaboration practices) and longer-term, strategic shifts (e.g., evolving governance models and participation norms) within OSS communities. These two directions follow prior socio-technical systems research emphasizing that sustainable change unfolds through the interplay of micro-level practices and macro-level structures that co-evolve over time \cite{trist1981evolution}. In the following section, we present our findings, organized into four categories structured according to STS components: (1) Software Development Practices (Tasks), (2) Documentation and Accessibility (Technology), (3) Community Engagement and Collaboration (People), and (4) Governance Practices (Structures).

\begin{table*}[ht]
\centering
\caption{Impact on Software Development Practices (Task)}
\label{tab:tech_impact}
\begin{adjustbox}{max width=\textwidth}
\begin{tabular}{P{3.5cm}P{3.5cm}P{3.5cm}P{3.5cm}P{3.5cm}}
\toprule\toprule
\textbf{Theme} & \textbf{Enhances} & \textbf{Obsolesces} & \textbf{Retrieves} & \textbf{Reverses} \\
\midrule\midrule

AI-Enhanced Testing Rigor &
Iteration speed and test coverage: AI accelerates development by generating test scaffolds, suggesting edge cases, and enabling rigorous testing workflows. &
TDD and manual test scaffolding: Traditional reliance on hand-written tests, like Test-Driven Development, is replaced by automated pipelines. &
Ad hoc experimentation culture: Contributors can prototype quickly, relying on AI to generate test prototypes. &
False assurance and blurred accountability: Over-reliance on AI tests may lead to unexamined approvals, missed edge cases, and weakened quality. \\
\midrule
AI-Enhanced Code Quality Tenets &
Baseline code quality: AI tools like Copilot ensure cleaner, readable code by flagging poor patterns and bugs across experience levels. &
Boilerplate coding and superficial reviews: Repetitive tasks like scaffolding and shallow style checks are offloaded to AI. &
Unix craftsmanship mindset: Revives simplicity, modularity, and maintainability principles. &
Vulnerability and false confidence: Blind reliance on AI code may introduce fragile or vulnerable code, approved without scrutiny. \\

\midrule
AI-Driven Technical Debt Management &
Technical debt awareness: AI tools scan codebases to identify outdated or redundant code, aiding remediation. &
Manual code audits: Periodic reviews and manual debt assessments are replaced by AI automation. &
Regular inspection and cleanup: Reinforces norms like removing dead code and enhancing codebase structure. &
Shallow fixes: Contributors may address flagged issues superficially, neglecting root causes and long-term design quality. \\
\midrule
AI-Augmented Code Review and Feedback &
Review efficiency and velocity: AI tools assist OSS reviews by generating comments and identifying issues, speeding up feedback. &
Long feedback loops: Reviewers avoid minor style checks, and contributors get faster feedback. &
In-depth design discussions: AI frees contributors to focus on architectural trade-offs and mentorship. &
Loss of community interaction: Generic AI comments may make reviews transactional, reducing human engagement and community bonds. \\

\midrule
AI-Automated Maintenance and Workflow Management &
Maintenance tasks: AI automates refactoring, formatting, and documentation updates, reducing maintainer fatigue. &
Low-level tedious work: AI alleviates minor bug fixes and code hygiene tasks. &
Fast self-healing OSS workflows: Enables rapid issue fixes, supporting "release early, release often." &
Prioritization loss: AI-generated commits may bury important issues, reducing maintainer visibility. \\
\bottomrule\bottomrule
\end{tabular}
\end{adjustbox}
\end{table*}

\textbf{Impact on Software Development Practices (Tasks):} GenAI may drive OSS toward higher quality by enhancing iteration speed, expanding test coverage, and enforcing rigorous testing workflows, while improving code quality through tools like Copilot that flag poor patterns and bugs, accommodating contributors of varying expertise levels within OSS collaborations. 
It may also increase visibility into technical debt and boost review efficiency through automated code analysis and feedback, easing maintainer workload and helping prioritize remediation efforts.
However, this progress comes with risks that can be reversed. Over-reliance on GenAI may introduce false assurance and blurred accountability, weaken quality by overlooking untested edge cases, and produce vulnerable code. In the meantime, manual testing, deep code audits, and hands-on contributor checks could be obsoleted, potentially leading to shallow fixes, neglected long-term design quality, and reduced human interaction in OSS communities. Additionally, for maintaining repositories, GenAI might prioritize less critical issues over important ones, impacting maintainer visibility on important issues. A detailed spectrum of these potential impacts is cataloged in Table~\ref{tab:tech_impact}, offering a framework for further exploration.

For instance, GenAI accelerates development cycles by automatically generating testing frameworks \cite{raghi2024software}, enabling robust contributions even from developers with limited coding expertise. These capabilities also render traditional scaffolding, manual test creation, and human-enforced style reviews obsolete, reducing opportunities for hands-on learning and critical reflection. GenAI then retrieves earlier OSS norms, such as ad hoc experimentation, where contributors can prototype quickly, relying on GenAI to generate test prototypes. Yet, this transformation may also reverse into false assurance and blurred accountability, where over-reliance on GenAI leads to the approval of shallow or low-quality code, potentially compromising the long-term integrity of software.

As these changes unfold, new tensions emerge between faster iteration and thoughtful review, between automation and craftsmanship, and between efficiency and accountability. For example, while AI can accelerate testing and refactoring, it can also reduce the depth of human code review and collective reflection.

\begin{bubble}{GenAI-augmented development in OSS}
streamlines iteration, testing, and maintenance in OSS, while potentially displacing human-centered practices like deep review, manual testing, and collaborative design. As GenAI accelerates development workflows, sustaining code integrity and shared understanding may depend on preserving opportunities for critical review, mentorship, and collective reflection within increasingly automated processes.
\end{bubble}

\begin{table*}[ht]
\centering
\caption{Impact on Documentation and Accessibility (Technology)}
\label{tab:ai_doc_impact}
\begin{adjustbox}{max width=\textwidth}
\begin{tabular}{P{3.5cm}P{3.5cm}P{3.5cm}P{3.5cm}P{3.5cm}}
\toprule\toprule
\textbf{Theme} & \textbf{Enhances} & \textbf{Obsolesces} & \textbf{Retrieves} & \textbf{Reverses} \\
\midrule\midrule

AI-Supported Code Comprehension and Community Contextual Learning &
Quick comprehension: AI tools enable rapid understanding of unfamiliar OSS codebases and workflows via contextual explanations and Q\&A. &
Exhaustive documentation reliance: Reduces the need to study lengthy READMEs, wikis, or discussion threads. &
Mentorship and contextual learning: Provides guidance to newcomers, reducing mentor burden and fostering inclusive onboarding. &
Misleading guidance and fragmented understanding: Over-reliance on AI may lead to vague or incorrect explanations, reducing documentation quality. \\
\midrule
AI-Driven Documentation Maintenance &
Documentation maintenance: AI syncs OSS documentation with code changes, easing updates in fast-moving projects. &
Manual documentation upkeep: Reduces the burden of revising documentation manually, ensuring alignment with code. &
Open contribution philosophy: Simplifies documentation, encouraging collaborative updates and broader participation. &
Silent divergence: AI-generated documentation may miss critical context, causing drift between code and its meaning. \\

\midrule
AI-Enhanced Code Explanation Practices &
Code clarity: AI tools encourage clearer comments, docstrings, and commit messages in OSS projects. &
Manual comment quality enforcement: AI flags unclear or missing explanations, reducing style-based rejections. &
Inline documentation culture: Revives context-rich, meaningful comments explaining code intent. &
Loss of human explanatory skills: Over-reliance on AI-generated summaries may weaken contributors’ ability to articulate code reasoning. \\
\midrule
AI-Enabled Multilingual Accessibility &
Multilingual accessibility: GenAI translation tools make OSS documentation accessible to non-English speakers. &
English-only defaults and manual translation: Reduces reliance on English and human translators with dynamic language support. &
Globalization ideals: Enables global, inclusive OSS communities with fewer linguistic barriers. &
Translation-induced errors: AI translations may introduce errors or awkward phrasing, misaligning technical or community-specific terms. \\
\bottomrule\bottomrule
\end{tabular}
\end{adjustbox}
\end{table*}

\textbf{Impact on Documentation and Accessibility (Technology):}  Within the \textit{Technology} dimension, our discussion focuses on the socio-technical artifacts that support contributors’ development tasks, such as documentation systems and accessibility infrastructures (e.g, translation services). These technologies directly shape who can participate, how contributors can contribute, and essential problem-solving tools for accomplishing OSS work.

GenAI is poised to revolutionize documentation and accessibility in OSS, cultivating a more inclusive and efficient ecosystem. GenAI may enhance comprehension by helping contributors interpret complex codebases, rendering unfamiliar OSS projects accessible to newcomers through contextual explanations and intuitive workflows. Additionally, multilingual accessibility may thrive as GenAI translation tools take down language barriers, enabling non-English speakers to contribute seamlessly to OSS documentation. However, the reliance on exhaustive, human-crafted documentation may diminish, potentially giving way to generic, standardized alternatives that lack depth and nuance. Furthermore, excessive reliance on automated documentation and translations risks weakening contributors’ capacity to articulate code reasoning and eroding their skills in crafting documentation tailored to community-specific nuances over time.

Table~\ref{tab:ai_doc_impact} details how GenAI introduces significant shifts in the enactment of documentation and accessibility practices. For instance, GenAI-powered code comprehension tools, or even non-fine-tuned, generic LLM models, can potentially enhance onboarding and participation by generating contextual explanations and enabling rapid familiarization with unfamiliar codebases and workflows \cite{tan2025revolutionizing}. This reduces reliance on traditional, exhaustive documentation artifacts such as lengthy, complex README files, CONTRIBUTING.md files, or discussion threads. These tools also retrieve mentorship and contextual learning traditions by offering targeted guidance that alleviates the cognitive and social burden of onboarding. This trend aligns with \citet{jackson2024creativity}, who note the obsolescence of traditional mentoring practices when discussing individual software development. However, such tools also risk reverting to misleading guidance or fragmented understanding, particularly when AI-generated explanations are vague, incorrect, or misaligned with evolving project norms.

Meanwhile, tensions are emerging between accessibility and authenticity, comprehension and understanding, and automation and authorship. For example, while GenAI expands who can access and understand OSS through translation and automated explanations, it might simultaneously dilute the depth of human-authored reasoning and community-specific context.

\begin{bubble}{AI-enhanced documentation and accessibility}
lower entry barriers, foster inclusivity in OSS through translation and contextual code explanations, but also risk diminishing human-authored nuance and weakening contributors’ ability to comprehend, reason about, and communicate design intent. Balancing automation with practices that preserve contextual depth and shared understanding may be key to sustaining participation and long-term maintainability.
\end{bubble}

\begin{table*}[ht]
\centering
\caption{Impact on Community Engagement and Collaboration (People)}
\label{tab:ai_engagement_impact}
\begin{adjustbox}{max width=\textwidth}
\begin{tabular}{P{3.5cm}P{3.5cm}P{3.5cm}P{3.5cm}P{3.5cm}}
\toprule\toprule
\textbf{Theme} & \textbf{Enhances} & \textbf{Obsolesces} & \textbf{Retrieves} & \textbf{Reverses} \\
\midrule\midrule

AI-Supported Contributor Retention/Engagement Management &
Retention awareness: AI automates onboarding, offboarding, and check-ins, maintaining consistent OSS contributor engagement. &
Manual retention efforts: Personalized onboarding and engagement tracking are replaced by AI-driven workflows. &
Newcomer community care: Revives OSS values of inclusive, proactive support for contributors. &
Trust issues and engagement erosion: Inaccurate or impersonal AI messages may confuse contributors, eroding trust and participation. \\
\midrule
AI-Facilitated Cross-Lingual Collaboration &
Cross-lingual communication: AI translation tools enable seamless collaboration across language barriers in OSS. &
English-dominated gatekeeping: AI reduces the need for English fluency or human translators. &
Multilingual participation: Revives OSS goals of global, localized knowledge access. &
Cultural miscommunication: AI mistranslations or awkward phrases may reduce credibility and cause confusion. \\

\midrule
AI-Enhanced Respectful, Inclusive Communication &
Community tone: AI identifies toxic language and promotes respectful, inclusive OSS communication. &
Manual moderation: AI reduces the need for human moderators to monitor tone. &
Code of Conduct ideals: Reinforces inclusive, welcoming OSS spaces for diverse contributors. &
Over-filtering and suppressed dialogue: AI’s excessive moderation may silence valid perspectives, weakening trust. \\
\midrule

AI-Augmented Learning and Mentorship &
24/7 knowledge access: AI acts as an always-available mentor, helping OSS contributors navigate issues and workflows. &
Traditional mentoring: Fixed-time office hours and formal mentoring are less necessary. &
Peer-driven learning: Revives informal collaboration and mentorship in OSS. &
Shallow learning and silos: Over-reliance on AI may lead to copy-paste habits and skipped foundational understanding. \\

\midrule
AI-Supported Interdisciplinary Collaboration &
Interdisciplinary collaboration: AI provides tailored explanations, easing cross-expertise collaboration in OSS. &
Long feedback loops: AI reduces delays from domain misunderstandings. &
Inclusive collaboration: Revives OSS’s vision of diverse, non-specialist contributions. &
Siloed understanding: Over-reliance on AI translations may hinder direct engagement, risking disconnected teams. \\
\bottomrule\bottomrule
\end{tabular}
\end{adjustbox}
\end{table*}

\textbf{Impact on Community Engagement and Collaboration (People):}
AI-supported retention strategies may raise awareness, streamline onboarding, and maintain consistent contributor engagement, replacing manual efforts with efficient, AI-driven workflows. Cross-lingual collaboration may flourish as GenAI translation tools enable seamless communication across language barriers, thereby revitalizing multilingual participation and localized knowledge within OSS~\cite{Botto-Tobar2025AI}. Guarded, respectful communication may supplant toxic language, promoting inclusive dialogue. At the same time, traditional mentoring may soon be complemented by 24/7 knowledge access from a specialized chatbot, acting as an always-available mentor to guide contributors through issues and workflows, and rekindling individualized mentoring programs' goals \cite{bagai2024designing}. This could be valuable for contributors from minoritized groups, where the scarcity of senior mentors with shared backgrounds often leads to mentor overload and burnout among available representatives \cite{jacobs2024mentorship}. In such cases, an AI mentor could provide consistent support without overburdening human mentors from underrepresented communities. The reliance on AI, however, may erode trust through impersonal messages and overly filtered communication, which may silence valid but critical discussion comments. English-dominated gatekeeping \cite{feng2023state} and traditional mentoring could also fade, with GenAI-created communications potentially leading to cultural misunderstandings and weakened human connections. Over time, such loss of interpersonal connection and mutual trust may diminish contributors’ sense of belonging and relatedness. As contributors feel less connected to peers and projects, disengagement may grow, potentially creating a cycle of declining activity and community vitality \cite{trinkenreich2024investigating}. Cultural misunderstandings are expected since LLMs are similar to people from WEIRD societies in cognitive psychological tasks \cite{atari2023which} and hence might not be a priori well-prepared for communication from other societies.

As detailed in Table~\ref{tab:ai_engagement_impact}, the embedding of GenAI in community workflows reshapes how contributors engage, collaborate, and support one another. For example, mentoring plays a crucial role in software development, especially in OSS, where almost every contributor has implicitly or explicitly mentored or has been mentored in their OSS contribution journey \cite{feng2024guiding}. AI-augmented learning tools increase access to domain knowledge by acting as on-demand mentors. This shift enhances contributor self-sufficiency and obsoletes fixed-time office hours and traditional mentoring formats. While such tools facilitate informal, self-paced learning, they may also lead to shallow learning experiences: contributors might adopt copy-paste habits or skip foundational understanding when mentorship becomes solely transactional.

These evolving practices might reveal tensions between inclusivity and authenticity, guidance and autonomy, and connection and efficiency. While GenAI broadens participation through accessible mentorship and multilingual communication, it could also distance contributors from one another, weakening interpersonal trust and shared community identity.

\begin{bubble}{AI-supported retention and communication}
enhance onboarding, multilingual collaboration, and inclusive communication in OSS, but risk eroding trust, cultural nuance, and authentic human connection. As AI mediates more interactions, communities must balance scalable, automated mentorship and translation with the preservation of authentic dialogue, critical feedback, and mutual learning that sustain OSS participation.
\end{bubble}

\begin{table*}[ht]
\centering
\caption{Impact on Governance Practices (Structures)}
\label{tab:ai_governance_impact}
\begin{adjustbox}{max width=\textwidth}
\begin{tabular}{P{3.5cm}P{3.5cm}P{3.5cm}P{3.5cm}P{3.5cm}}
\toprule\toprule
\textbf{Theme} & \textbf{Enhances} & \textbf{Obsolesces} & \textbf{Retrieves} & \textbf{Reverses} \\
\midrule\midrule
AI-Supported Governance Docs &
Governance drafting: AI aids in drafting comprehensive governance documents like Codes of Conduct and contribution guidelines. &
Delays in governance documentation: Reduces delays in creating foundational documents, especially for new OSS projects. &
Governance scaffolding: Establishes norms that support contributor participation in smaller projects. &
Governance as placeholder: AI-generated documents may become superficial, rarely read or enforced. \\
\midrule
AI-Assisted Licensing and Legal Guidance &
License support: AI validates license compatibility and provides 24/7 Q\&A, improving legal awareness. &
Compliance checks: Eliminates delays in legal reviews, offering instant guidance for licensing issues. &
--- & 
Compliance noise: AI may produce inaccurate advice or false positives. \\
\midrule
AI-Driven Policy Enforcement &
Policy enforcement: AI flags violations of project policies, ensuring consistent enforcement in OSS. &
Human-only moderation: Reduces moderator burden and supports newcomers in reporting issues. &
Inclusive contribution: Reinforces safe, accountable OSS environments for all contributors. &
Surveillance discomfort: Overuse of AI may make contributors feel monitored, eroding trust. \\
\bottomrule\bottomrule
\end{tabular}
\end{adjustbox}
\end{table*}

\textbf{Impact on Governance Practices (Structures):} The organization component in STS reflects the formal structures, norms, and governance mechanisms that coordinate collective activity in OSS communities.  AI-supported governance documents may streamline the drafting process, enhancing the comprehensiveness of codes of conduct and contribution guidelines. Meanwhile, AI-assisted licensing provides instant support to maintainers by integrating license validation and compatibility checks directly into project workflows and CI/CD pipelines, ensuring continuous compliance and raising awareness through 24/7 Q\&A assistance. Moreover, AI-driven policy enforcement may strengthen project policies, ensuring consistent adherence across OSS ecosystems. However, GenAI may introduce compliance noise by producing inaccurate advice or suggesting incorrect licensing configurations. Additionally, Human-only moderation might fade, replaced by GenAI surveillance, which could potentially discomfort contributors who feel monitored, thereby eroding trust.

Table~\ref{tab:ai_governance_impact} outlines the transformative effects of AI-enhanced tools on governance practices within OSS ecosystems. For example, AI-supported tools for governance documentation help automate the creation of foundational documents, such as Codes of Conduct and contribution guidelines \cite{cobos2025bot, poth2024considerations}. These tools enhance procedural efficiency and reduce delays in documentation, particularly for new projects where maintainers have no experience in governance documentation. By lowering the barrier to governance articulation, GenAI retrieves earlier OSS scaffolding norms that encourage contributor participation in governance, especially within smaller or loosely organized communities. However, this shift may also reverse into superficiality: AI-generated governance text risks becoming a placeholder—present in form but rarely read, internalized, or enforced by the community.

While GenAI strengthens governance consistency and reduces human workload, tensions arise between procedural efficiency and community legitimacy, automation and accountability, and enforcement and trust. As oversight becomes increasingly automated, maintaining transparent, participatory governance becomes more challenging.

\begin{bubble}{AI-enhanced governance practices}
streamline documentation, licensing, and policy enforcement in OSS, but risk introducing compliance noise and eroding trust through automated moderation and surveillance. Future governance models must balance procedural efficiency with participatory oversight, ensuring that automation strengthens, not substitutes, transparency, accountability, and contributor trust.
\end{bubble}

\section{Discussion and Research Directions}
\label{sec:discussion}

Our tetrad analysis highlights how GenAI reshapes human and social dynamics in OSS, with each anticipated impact pointing to a potential research direction. Table~\ref{tab:ai_traceability} maps the impacts surfaced through the Tetrad analysis to potential areas for future research. Building on our scenarios that examine the interplay between micro-level practices and macro-level structures, we propose a two-stage research agenda: near-term shifts, reflecting immediate adaptations to AI integration, and far-reaching transformations, representing longer-term structural and cultural evolution in OSS. The following subsections elaborate on these research directions.

\begin{table*}[ht]
\centering
\caption{Traceability Between GenAI Impacts and Proposed Research Directions.}
\label{tab:ai_traceability}
\resizebox{\columnwidth}{!}{\begin{tabular}{lll}
\hline\hline 
\multicolumn{1}{c}{\textbf{Direction}}                                                                                     & \multicolumn{1}{c}{\textbf{Rooted Impacts}}                                                                                                                                                                                                                                                                                                                                                                                                                                                                                                                         & \multicolumn{1}{c}{\textbf{Conceptual Link}}                                                                                                                                                                          \\ \hline\hline
\multicolumn{3}{c}{\textbf{Near-Term Shifts in OSS Practices}}                                                                                                                                                                                                                                                                                                                                                                                                                                                                                                                                                                                                                                                                                                                                                                                                                                                           \\ \hline\hline

\begin{tabular}[c]{@{}l@{}}Hybrid \\ Contribution \\ Model\end{tabular}                                                    & \begin{tabular}[c]{@{}l@{}}AI-Enhanced Testing Rigor (Task)\\ AI-Enhanced Code Quality Tenets (Task)\\ AI-Augmented Code Review and Feedback (Task)\\ AI-Automated Maintenance (Task)\\ AI-Supported Interdisciplinary Collaboration (People)\end{tabular}                                                                                                                                                                                                                                                                                                          & \begin{tabular}[c]{@{}l@{}}These impacts reveal that AI \\ automates core OSS activities, \\ prompting new hybrid task \\ allocation between humans \\ and AI.\end{tabular}                                           \\ \hline
\begin{tabular}[c]{@{}l@{}}Rethinking \\ Governance \\ for Human–AI \\ Collaboration\end{tabular}                          & \begin{tabular}[c]{@{}l@{}}AI-Enhanced Testing Rigor (Task)\\ AI-Automated Maintenance (Task)\\ AI-Supported Governance Docs (Structures)\\ AI-Assisted Licensing and -\\ - Legal Guidance (Structures)\end{tabular}                                                                                                                                                                                                                                                                                                                                            & \begin{tabular}[c]{@{}l@{}}These impacts expose blurred \\ accountability and evolving \\ authorship boundaries, \\ motivating research on \\ redefining governance roles.\end{tabular}                               \\ \hline

\begin{tabular}[c]{@{}l@{}}Building \\ Trust in \\ AI-Driven \\ OSS \\ Governance\end{tabular}                             & \begin{tabular}[c]{@{}l@{}}AI-Enabled Multilingual Accessibility (Technology)\\ AI-Supported Contributor Retention (People)\\ AI-Facilitated Cross-Lingual Collaboration (People)\\ AI-Enhanced Respectful, Inclusive Communication (People)\\ AI-Supported Governance Docs (Structures)\\ AI-Driven Policy Enforcement (Structures)\end{tabular}                                                                                                                                                                                                               & \begin{tabular}[c]{@{}l@{}}These impacts show that \\ automated moderation and \\ governance affect trust and \\ fairness, motivating studies \\ on maintaining legitimacy \\ in AI-assisted governance.\end{tabular} \\ \hline
\begin{tabular}[c]{@{}l@{}}Revisiting \\ Code Quality \\ and Technical \\ Debt\end{tabular}                                & \begin{tabular}[c]{@{}l@{}}AI-Enhanced Testing Rigor (Task)\\ AI-Enhanced Code Quality Tenets (Task)\\ AI-Automated Maintenance (Task)\\ AI-Driven Technical Debt Management (Task)\\ AI-Driven Documentation Maintenance (Technology)\end{tabular}                                                                                                                                                                                                                                                                                                                                                              & \begin{tabular}[c]{@{}l@{}}These impacts identify new, \\ invisible forms of AI-induced \\ technical debt, motivating research \\ on rethinking code quality metrics.\end{tabular}                                    \\ \hline

\begin{tabular}[c]{@{}l@{}}Documentation \\ and \\ Accessibility\end{tabular}                                              & \begin{tabular}[c]{@{}l@{}}AI-Supported Code Comprehension (Technology)\\ AI-Driven Documentation Maintenance (Technology)\\ AI-Enhanced Code Explanation Practices (Technology)\\ AI-Enabled Multilingual Accessibility (Technology)\\ AI-Facilitated Cross-Lingual Collaboration (People)\end{tabular}                                                                                                                                                                                                                                                        & \begin{tabular}[c]{@{}l@{}}These impacts reveal both benefits \\ and pitfalls of automated \\ documentation, motivating inquiry \\ into sustaining documentation \\ reliability and inclusivity.\end{tabular}         \\ \hline\hline
\multicolumn{3}{c}{\textbf{Far-Reaching Transformations in OSS Development}}                                                                                                                                                                                                                                                                                                                                                                                                                                                                                                                                                                                                                                                                                                                                                                                                                                             \\ \hline\hline

\begin{tabular}[c]{@{}l@{}}Sustaining \\  Contributor\\  Skill Development\end{tabular}                                         & \begin{tabular}[c]{@{}l@{}}AI-Enhanced Code Quality Tenets (Task)\\  AI-Supported Code Comprehension (Technology)\\ AI-Enhanced Code Explanation Practices (Technology)\\ AI-Augmented Learning and Mentorship (People)\\ AI-Supported Interdisciplinary Collaboration (People)\end{tabular}                                                                                                                                                                                                                            & \begin{tabular}[c]{@{}l@{}}These impacts reveal how AI \\ reshapes developer learning and \\ craftsmanship, motivating \\ long-term research on sustained \\ expertise and knowledge \\ evolution.\end{tabular}       \\ \hline
\begin{tabular}[c]{@{}l@{}}Sustaining \\ Human \\ Participation \\ in \\ AI-Augmented \\ OSS \\ Collaboration\end{tabular} & \begin{tabular}[c]{@{}l@{}}AI-Augmented Code Review and Feedback (Task)\\ AI-Supported Code Comprehension (Technology)\\ AI-Enhanced Code Explanation Practices (Technology)\\ AI-Enabled Multilingual Accessibility (Technology)\\ AI-Supported Contributor Retention (People)\\ AI-Facilitated Cross-Lingual Collaboration (People)\\ AI-Enhanced Respectful, Inclusive Communication (People)\\ AI-Augmented Learning and Mentorship (People)\\ AI-Supported Interdisciplinary Collaboration (People)\\ AI-Driven Policy Enforcement (Structures)\end{tabular} & \begin{tabular}[c]{@{}l@{}}These impacts show that automation \\ may weaken trust and mentorship, \\ motivating research on sustaining \\ meaningful human participation in \\ AI-mediated OSS.\end{tabular}          \\ \hline

\begin{tabular}[c]{@{}l@{}}Clarifying the \\ Responsibilities \\ of GenAI Providers \\ in OSS\end{tabular}                 & \begin{tabular}[c]{@{}l@{}}AI-Automated Maintenance (Task)\\ AI-Enhanced Respectful, Inclusive Communication (People)\\ AI-Assisted Licensing and Legal Guidance (Structures)\\ AI-Driven Policy Enforcement (Structures)\end{tabular}                                                                                                                                                                                                                                                                                                                    & \begin{tabular}[c]{@{}l@{}}These impacts expose governance and \\ accountability gaps created by AI tools, \\ motivating research on defining GenAI \\ providers’ ethical and legal responsibilities.  \end{tabular}  \\ \hline\hline 
\end{tabular}}
\end{table*}

\subsection{Near-Term Shifts in OSS Practices.}

The near-term shifts outlined below capture the immediate adaptations now taking shape as OSS communities integrate GenAI into their workflows. These shifts involve rapid experimentation, evolving norms, and early frictions in human–AI collaboration. Because these changes are actively unfolding, they offer a timely opportunity for researchers to observe, model, and intervene before new practices and structures become entrenched.

\textbf{Collaboration: Hybrid Contribution Model.} 
As GenAI tools become increasingly embedded in software development workflows, OSS communities are likely entering an era of hybrid collaboration between human contributors and AI systems. 
However, the absence of clearly defined practices for integrating AI-generated artifacts into OSS ecosystems presents a growing challenge. 
Without standardized guidance, communities face uncertainty regarding authorship, accountability, and reproducibility, which raises concerns such as the quality of software, the trustworthiness of collaborations, and the inclusiveness of participation.
To address this gap, research needs to investigate: \textit{How should OSS communities design and operationalize hybrid contribution workflows that integrate AI-generated and human-authored artifacts?}
This question aims to identify contribution models that not only maintain software quality but also enhance OSS productivity and innovation, while preserving the core values of openness, collaboration, and inclusivity.
Future research could unpack this direction by investigating: (1) How standardized attribution formats should be designed to clearly distinguish between AI-generated and human-authored code, documentation, and other artifacts?
(2) What kinds of human oversight protocols are needed to ensure that AI-assisted contributions maintain high standards of code quality, minimize bias, and meet ethical and reliability expectations?
(3) What mechanisms are required to make prompt histories transparent, version-controlled, and aligned with OSS norms for reproducibility and auditability?

\textbf{Governance: Rethinking Governance for Human-AI Collaboration.} 
When OSS communities move toward human-AI collaboration, existing governance structures must evolve to support these emerging models.
Yet, current governance practices were not designed to address such complexities.
Without reexamining governance practices, OSS communities face uncertainty around authorship, licensing, accountability, and liability, necessitating answers to: 
\textit{What governance practices are needed to manage human–AI collaboration in OSS?}
Answers to this question could help identify governance mechanisms that OSS communities can adopt to manage risk, clarify contributor roles and responsibilities, and support transparent decision-making.
To do so, future research could explore:
(1) How should licensing and authorship be attributed in code generated collaboratively by humans and AI?
(2) How can OSS communities ensure transparency and contestability in AI-generated governance decisions?
(3) How should OSS communities negotiate copyright and liability when AI-generated artifacts lead to downstream harm or conflict?

\textbf{Fairness \& Trust: Building Trust in AI-Driven OSS Governance.} 
OSS communities are beginning to integrate AI into governance workflows, such as moderation \cite{hsieh2023nip} and code of conduct enforcement \cite{cobos2025bot}.
There is a growing concern that AI-driven governance may be perceived as arbitrary, biased, or opaque, especially when it lacks context or transparency.
These risks are particularly acute in already diverse and globally distributed projects, where issues of bias, power imbalance, and unequal participation are well documented \cite{feng2023state}.
This creates the need to explore: \textit{How OSS communities can build and maintain trust in AI-assisted governance processes?}
Answering this question could guide the design of trustworthy and culturally sensitive AI governance tools that strengthen contributor confidence and participation.
Future research could explore
(1) How do contributors perceive the legitimacy and transparency of AI-assisted governance decisions, and what forms of human oversight increase trust?
(2) How can fairness perceptions be maintained when governance decisions are partially or fully supported by AI systems, especially across geographically and culturally diverse contributors?
(3) What mechanisms can detect and mitigate bias in AI-supported multilingual communication and moderation?

\textbf{Sustainability: Revisiting Code Quality and Technical Debt.} 
AI-powered tools are increasingly shaping how software is written, reviewed, and maintained.
However, they may introduce new forms of long-term technical debt --- LLM smells---that remain poorly understood and are difficult to detect with current tooling.
Such technical debt can accumulate silently, degrading maintainability, reducing performance, or embedding biased logic into critical systems, which can cause harm that may only surface much later in the software lifecycle. While technical debt is a pervasive concern across all software projects, it poses particular risks in OSS due to decentralized governance, volunteer-driven maintenance, and uneven contributor expertise \cite{haddad2020technical}.
To explore these emerging concerns, one can ask: \textit{How can OSS communities address new forms of technical debt introduced by LLM-assisted development?}
Answering this question is timely and important, as it can help OSS communities proactively identify emerging risks and develop new practices and tools to sustain code quality, maintainability, and trust in an era of hybrid human–AI collaboration.
To unpack this question, future research could start with directions, such as
(1) What types of code patterns or anti-patterns emerge uniquely from LLM-generated code, and how do they differ from traditional code smells?
(2) What tools, metrics, or review processes are needed to identify and mitigate these LLM smells before they accumulate as long-term debt?

\textbf{Accessibility \& Inclusion: Documentation and Accessibility.} 
Well-structured, comprehensible documentation directly determines who can access and contribute within OSS communities \cite{fronchetti2023contributing}. GenAI has demonstrated capabilities in automating repetitive and routine documentation tasks in OSS projects \cite{luo2024repoagent}, such as syncing documentation with code or generating contextual explanations for contributors, and improving accessibility by lowering language and expertise barriers.
However, as GenAI makes it easier to generate abundant documentation, there is a growing risk that meaningful context, including design rationale, architectural decisions, and developer intent, may become buried in noise or be omitted entirely.
Losing this context can weaken the accessibility and maintainability of OSS projects, making it harder for contributors to understand why certain decisions were made or how systems have evolved. 
This raises the question: \textit{What are the impacts of GenAI-driven documentation automation on long-term maintainability?}
The answer to this question can inform the design of GenAI-powered documentation tools and workflows that produce efficient, effective, and context-rich documentation.
Research could unpack: 
(1) What documentation formats or workflows best preserve project knowledge over time while integrating AI support?
(2) How do contributors perceive and engage with AI-generated documentation during onboarding or code-comprehension tasks?

\subsection{Far-Reaching Transformations in OSS Development.}

While near-term shifts describe immediate adaptations, they also initiate longer-term transformations. As these adaptations solidify into shared norms and governance structures, they gradually reshape contributor roles and community dynamics. Researchers can already begin tracing early indicators of these transformations as GenAI becomes more deeply embedded in OSS ecosystems.

\textbf{Contributor Growth: Sustaining Contributor Skill Development.}
It is crucial to investigate how GenAI affects not only productivity but also the development and maintenance of OSS-specific skills and community practices over a longer timeframe.
While GenAI tools may boost short-term efficiency, their long-term effects on competencies, such as code comprehension, debugging, and collaborative problem-solving, remain poorly understood. 
If contributors’ reliance on GenAI leads to skill atrophy or bypasses traditional mentorship dynamics, OSS communities may struggle to cultivate expertise, transfer knowledge, or sustain peer learning over time. 
These uncertainties prompt the question:
\textit{How does long-term GenAI integration affect contributor skill development, mentorship, and sustainability in OSS communities?}
This question is crucial for understanding how OSS communities, tool builders, mentoring programs, and researchers can design workflows, mentorship models, and learning resources that preserve critical skills, foster expertise, and ensure sustainable contributor development.
To explore this direction, researchers could explore:
(1) How does prolonged use of GenAI tools affect contributors’ ability to comprehend, debug, and maintain OSS codebases?
(2) What are the long-term effects of GenAI on traditional mentorship structures, onboarding, and peer learning in OSS?

\textbf{Collaboration: Sustaining Human Participation in AI-Augmented OSS Collaboration.} 
GenAI-powered bots are becoming increasingly prevalent in OSS workflows \cite{cobos2025bot, tan2025revolutionizing}.
The increasing reliance on AI to handle coding, reviewing, and maintenance tasks risks marginalizing or replacing human contributors.
When AI mediates both code and collaboration, the opportunity for human connection, mentorship, and informal learning may be significantly reduced, potentially weakening the social bonds that sustain OSS communities over time. We need to ask:
\textit{What is the long-term impact of GenAI-driven automation on contributors' participation?}
Answering this question may help support sustainable participation in OSS by identifying how collaboration models must evolve in response to automation and guiding communities in designing inclusive pathways for engagement in the GenAI era.
To explore this direction, future research could ask:
(1) How does GenAI-Driven automation reshape contributors' identities, motivations, and sense of belonging in OSS?
(2) In what ways do automation-driven shifts in collaborations affect retention, recognition, and acknowledgment?

\textbf{Governance: Clarifying the Responsibilities of GenAI Providers in OSS Ecosystems.}
GenAI introduces new dependencies into the community infrastructure.
There are no clear frameworks defining the responsibilities of GenAI providers.
Without effective accountability mechanisms, projects may inadvertently become overly dependent on AI-powered tools. GenAI tools risk reinforcing historical biases and undermining OSS values such as transparency, openness, and collective governance. 
It is important to explore
\textit{What responsibilities should GenAI providers have toward OSS communities, and how can these responsibilities be formalized and enforced?}
Answering this question could guide the development of accountability frameworks that define how GenAI providers should disclose training data, address harmful outputs, engage with affected communities, and report system-level changes.
To investigate this direction, research could start to investigate:
(1) What types of training data disclosures are necessary to ensure transparency when GenAI tools are used in OSS development?
(2) How can OSS communities hold GenAI providers accountable for biased or harmful outputs that affect contributor participation or project direction?

\textbf{Research Direction Prioritization.}
While all discussed research directions are important, they vary in immediacy and dependency. The near-term directions, such as rethinking hybrid contribution models, documentation practices, and governance mechanisms for human–AI collaboration, represent foundational inquiries that can be empirically studied today as GenAI tools are actively integrated into OSS workflows. Insights from these efforts may, in turn, inform longer-term transformations concerning sustainability, contributor participation, and ethical responsibility. In this sense, the two-stage agenda we propose is not hierarchical but iterative: progress on near-term adaptations enables a deeper understanding of long-term socio-technical change.

\subsection{Practical Implications for OSS Practitioners.}

\begin{table*}[ht]
\centering
\caption{Practitioner Reflection Guide}
\label{tab:practioner}
\begin{adjustbox}{max width=\textwidth}
\renewcommand{\arraystretch}{1.25}
\begin{tabular}{
    >{\raggedright\arraybackslash}p{2.8cm}  
    >{\raggedright\arraybackslash}p{2.5cm}  
    >{\raggedright\arraybackslash}p{3cm}    
    >{\raggedright\arraybackslash}p{7.5cm}  
}
\hline\hline
\textbf{STS Components} & 
\textbf{Role} & 
\textbf{Signals to Watch} & 
\textbf{Potential Ways to Use} \\
\hline\hline

\textbf{Development Practices (Tasks)} &
Maintainers, Developers &
Over-automation, shallow fixes, blurred accountability &
Use Table~\ref{tab:tech_impact} as a \textit{checklist during sprint retrospectives}, prompting questions such as “Which of our workflows are moving toward speed over reflection?” \\
\hline
\textbf{Documentation \& Accessibility (Technology)} &
Technical Writers, Onboarding Leads &
Loss of nuance, generic AI documentation &
Use Table~\ref{tab:ai_doc_impact} as a \textit{mapping tool during documentation reviews}: for each artifact (e.g., README, CONTRIBUTING.md, tutorials), identify which sections could be AI-assisted and note where a human interpretive layer is needed to preserve project-specific context. \\
\hline
\textbf{Community Engagement \& Collaboration (People)} &
Mentors, Community Managers &
Declining trust, impersonal mentoring, cultural miscommunication &
Use Table~\ref{tab:ai_engagement_impact} to \textit{guide discussions during monthly community or governance management meetings}, focusing on how GenAI-driven changes in mentoring, communication, and participation are shaping contributor engagement. \\
\hline
\textbf{Governance Practices (Structures)} &
Project Leaders, Foundation Stewards &
Compliance noise, excessive surveillance, legitimacy loss &
Use Table~\ref{tab:ai_governance_impact} as a \textit{governance-readiness checklist during policy or documentation reviews}: for each clause, identify which aspects could be automated and which require explicit human oversight to maintain legitimacy and trust. \\
\hline\hline
\end{tabular}
\end{adjustbox}
\end{table*}

While this paper primarily outlines research directions, it also offers practical avenues for OSS practitioners to act upon today. Table~\ref{tab:practioner} translates our scenarios into reflection guides that practitioners can use during retrospectives, documentation reviews, community meetings, or governance updates. For instance, when considering development practices, OSS leaders and maintainers can use our STS-guided framework to evaluate which workflows are drifting toward speed over reflection, identify where AI-assisted documentation requires human context, and discuss how GenAI affects mentoring and participation. Foundation stewards can apply the governance-readiness checklist to determine which policies can be automated and which demand human oversight.

Beyond suggestions, we encourage practitioners to actively interact with this article by leading discussions around the scenarios, using the tables as checklists of topics to explore, or adapting the questions to their own projects. Treating the table as a recurring discussion aid or initiative framework allows practitioners to anticipate unintended consequences of GenAI adoption and ensure that automation strengthens, rather than substitutes, the human values that sustain OSS collaboration.

\subsection{Limitation}

As with many studies, we acknowledge that our work has limitations. In this section, we discuss limitations inherent to our approach, including those related to our choice of conceptual lens, the perspectives represented by the author team, the thematic organization of impacts, and the potential future research directions.

\textbf{Conceptual approach.}
Our analysis uses McLuhan's Tetrad \cite{mcluhan} as an analytic lens within the  STS-guided framework to create interpretative scenarios of GenAI usage in OSS communities. We chose this framework, instead of other theories and models such as, Activity Theory \cite{feng2025domains}, Structuration Theory \cite{stones2017structuration}, or the Job Demands–Resources model \cite{feng2025modeling}, as it is particularly well-suited for early explorations of emerging technological disruptions, offering a balanced structure for connecting technological effects with human practices and organizational change \cite{leavitt2013applied, lyytinen2008explaining, ciriello2024emergence}.

\textbf{Author expertise.}
The scenarios presented in this paper reflect the collective perspectives of a multidisciplinary team of seven authors, which are likely shaped by our experiences and specialization. Other researchers with different lived experiences and viewpoints may have other interpretations of the impact of GenAI on OSS. The goal is that these very differences in perspectives may serve as catalysts for discussion that prompt researchers and practitioners to come together to envision alternate futures of OSS communities in the era of GenAI and work together to create responsible use of AI in OSS.

\textbf{Thematic grouping.}
We acknowledge that the thematic grouping used in our Tetrad tables reflects an inductive synthesis guided by the authors’ shared understanding of OSS socio-technical contexts. The groupings were collaboratively developed to illustrate representative impacts within each STS components, rather than to provide exhaustive or explanatory insights. Alternative categorizations of the identified effects are possible and may emerge under different analytical framings or disciplinary perspectives. Future empirical work could extend or refine these thematic structures to reflect evolving practices in OSS communities.

\textbf{Research directions.}
Our proposed research directions present a possible roadmap and are not meant to be exhaustive. Each direction was developed to extend the implications of the identified impacts into actionable areas for future study; however, other researchers may derive alternative directions depending on their theoretical orientation, disciplinary expertise, or empirical focus. We acknowledge this as an inherent limitation of translating conceptual insights into a research agenda, one that aggregates the broader limitations discussed above, including those related to our conceptual framing, author expertise, and thematic grouping. We encourage future work to expand, refine, or challenge these directions as the socio-technical dynamics of GenAI in OSS continue to evolve.

\subsection{Conclusion} 
Nobody knows exactly how OSS may evolve as GenAI becomes deeply and increasingly embedded in development practices. What is certain, however, is that the nature of contributions, collaboration, and community may continue to shift. Concerns persist that the rapid advancement of AI may disrupt the collaborative ethos that has long defined OSS. At the same time, GenAI offers opportunities to strengthen the ecosystem, enabling new forms of innovation, participation, and support.  Much of the tension surrounding these changes may stem not just from GenAI's specific impacts, but from the inherent challenge of technological disruption itself, which can be disorienting and conflict-inducing even when outcomes prove beneficial.

In this work, we have discussed both hopeful possibilities and emerging tensions introduced by GenAI into the OSS landscape. This paper is not a prediction, but a catalyst—a call to researchers, community leaders, contributors, tool builders, and practitioners—to recognize, support, and adapt to the socio-technical dynamics that are already unfolding. It charts a path not through forecasts but through framing, clarifying the forces at play, the questions that demand attention, and the choices communities must make as GenAI reshapes how we build and sustain OSS ecosystems.

\section*{Acknowledgments}
This work is partially supported by NSF grants 2303612, 2247929, 2303042, 2347311, and 2303043.

Sandia National Laboratories is a multi-mission laboratory managed and operated by National Technology \& Engineering Solutions of Sandia, LLC (NTESS), a wholly owned subsidiary of Honeywell International Inc., for the U.S. Department of Energy’s National Nuclear Security Administration (DOE/NNSA) under contract DE-NA0003525. This written work is authored by an employee of NTESS. The employee, not NTESS, owns the right, title and interest in and to the written work and is responsible for its contents. Any subjective views or opinions that might be expressed in the written work do not necessarily represent the views of the U.S. Government. The publisher acknowledges that the U.S. Government retains a non-exclusive, paid-up, irrevocable, world-wide license to publish or reproduce the published form of this written work or allow others to do so, for U.S. Government purposes. The DOE will provide public access to results of federally sponsored research in accordance with the DOE Public Access Plan.

\def\refname{REFERENCES}

\bibliographystyle{ACM-Reference-Format}
\bibliography{bib.bib}

\end{document}